\documentclass{emulateapj}
\usepackage{apjfonts}
\usepackage[]{natbib}
\usepackage{graphics}
     
\newcommand{\etal}{et al.}  
\newcommand{\per}{\ensuremath{^{-1}}}
\newcommand{\persq}{\ensuremath{^{-2}}}

\newcommand{\hal}{H\ensuremath{\alpha}}
\newcommand{\hbeta}{H\ensuremath{\beta}} 
\newcommand{\hst}{\emph{HST}}
\newcommand{\msun}{\ensuremath{M_{\odot}}}
\newcommand{\kms}{km s\ensuremath{^{-1}}}

\newcommand{\mgb}{Mg\ensuremath{b}}

\slugcomment{}
\shorttitle{NORMAL STELLAR DISK IN MALIN 1} 
\shortauthors{BARTH}

\begin{document} 

\title{A Normal Stellar Disk in the Galaxy Malin 1$^1$}

\author{Aaron J. Barth}

\affil{Department of Physics and Astronomy, 4129 Frederick
  Reines Hall, University of California, Irvine, CA 92697-4575;
  barth@uci.edu}

\begin{abstract}

Since its discovery, Malin~1 has been considered the prototype and
most extreme example of the class of giant low surface brightness disk
galaxies.  Examination of an archival \emph{Hubble Space Telescope}
$I$-band image reveals that Malin~1 contains a normal stellar disk
that was not previously recognized, having a central $I$-band surface
brightness of $\mu_0 = 20.1$ mag arcsec\persq\ and a scale length of
4.8 kpc.  Out to a radius of $\sim10$ kpc, the structure of Malin~1 is
that of a typical SB0/a galaxy.  The remarkably extended, faint outer
structure detected out to $r \approx 100$ kpc appears to be a
photometrically distinct component and not a simple extension of the
inner disk.  In terms of its disk scale length and central surface
brightness, Malin~1 was originally found to be a very remote outlier
relative to all other known disk galaxies.  The presence of a disk of
normal size and surface brightness in Malin~1 suggests that such
extreme outliers in disk properties probably do not exist, but
underscores the importance of the extended outer disk regions for a
full understanding of the structure and formation of spiral galaxies.

\end{abstract}


\section{Introduction}

The galaxy Malin 1 has remained a singularly unusual and puzzling
object for nearly two decades.  As described in the discovery paper by
\citet{bot87} and in later work by \citet{ib89}, it has the largest
radial extent of any known spiral galaxy, with low-surface brightness
emission extending out to $\sim100$ kpc, and its disk was found to
have an extrapolated central surface brightness of only $\mu_0(V)
\approx 25.5$ mag arcsec\persq, far fainter than any galaxy previously
known.  The exponential scale length of this disk was determined to be
$\sim70$ kpc.  Despite the faint surface brightness of its disk, Malin
1 is a massive galaxy with a total optical luminosity of $M_V \approx
-22.9$ mag \citep{pic97}.  Based on these properties, it is considered
the prototypical giant low surface brightness (LSB) galaxy.  Even in
comparison with other giant LSB galaxies discovered subsequently
\citep{bot90, spr93, spr95, bei99}, the properties of Malin 1 are
unusual; no other galaxy disk has been found with a surface brightness
and scale length that approach these extreme values.  Furthermore, it
is one of the most gas-rich galaxies known, with an \ion{H}{1} mass of
$\sim5\times10^{10}$ \msun\ \citep{pic97, mvm01}.  The \ion{H}{1}
rotation curve measured by \citet{pic97} reaches an asymptotic value
of $v_r = 190$ \kms\ and extends to radii as large as 110 kpc from the
galaxy center, implying a dynamical mass of order $10^{12}$ \msun\
although the disk inclination is uncertain and the velocity structure
shows some warping and noncircular motions.

The implications of this object for understanding the overall
population of disk galaxies are significant.  As emphasized by
\citet{bim97}, galaxies like Malin 1 would likely be underrepresented
in magnitude-limited galaxy surveys because of surface brightness
selection effects.  Even if such extreme giant LSB galaxies are rare,
which appears to be the case \citep[e.g.,][]{min04}, the properties of
Malin 1 pose interesting challenges to theories of galaxy formation.
Models of disk formation in cold dark matter halos are not easily able
to reproduce the properties of such enormously extended disks,
although scenarios have been proposed which might account for the
formation of Malin 1-type galaxies as rare density peaks in voids
\citep{hsw92} or by secular evolution and radial mixing in ordinary
disk galaxies \citep{nog01}.  As the most diffuse and gas-rich giant
LSB galaxy known, Malin 1 is also a particularly interesting
laboratory for understanding issues of disk stability and thresholds
for the onset of star formation \citep{ib89}.

This paper presents a reexamination of an archival \emph{Hubble Space
Telescope} (\hst) image of Malin 1, which reveals that the galaxy's
optical morphology is rather different from what had previously been
determined based on ground-based imaging.  It is shown that Malin 1
actually contains a fairly normal stellar disk typical of an
early-type barred spiral, and that based on the presence of this disk
Malin 1 does not satisfy the standard criteria for being classified as
an LSB galaxy despite its having the most extended low surface
brightness outer regions of any known disk galaxy.  This is not merely
a semantic distinction, as this new observation significantly
truncates the range of parameter space (in terms of central surface
brightness and scale length) populated by real galaxy disks by
removing an extremely aberrant data point from the observed galaxy
population.  A new optical spectrum of the galaxy is also used to
derive its bulge velocity dispersion and to reexamine the
classification of its active nucleus.

Malin 1 has a redshift of 0.0825 \citep{pic97}, corresponding to a
luminosity distance of $D_L = 370$ Mpc and a projected angular scale
of 1.53 kpc arcsec\per\ for a cosmology with $H_0 = 71$ km s\per\
Mpc\per, $\Omega_M = 0.27$, and $\Omega_\Lambda = 0.73$.  For
consistency, observed quantities taken from the literature and listed
in this paper have been rescaled to this cosmology.

\setcounter{footnote}{1}

\footnotetext{Based on observations made with the NASA/ESA Hubble Space
  Telescope, obtained from the Data Archive at the Space Telescope
  Science Institute, which is operated by the Association of
  Universities for Research in Astronomy, Inc., under NASA contract
  NAS 5-26555.}

\section{Archival Data and Observations}

Images of Malin 1 were obtained with the WFPC2 camera as part of \hst\
program GO-5946 (PI: Impey) on 1996 March 10.  The galaxy was observed
for one orbit each in the F300W ($U$-band) and F814W ($I$-band)
filters, with the galaxy placed on the WF3 CCD near the center of the
full WFPC2 field of view.  Each orbit was split into four individual
exposures for removal of cosmic-ray hits.  The F300W exposures were
read out with $2\times2$ pixel binning on the CCD, while the F814W
images were read out in the standard unbinned mode.  Total exposure
times were 2100 s in each filter.  This dataset was previously
analyzed by \citet{obi00}, who studied the properties of faint
galaxies in the same WFPC2 field as Malin 1, but they did not discuss
the structure of Malin 1 itself.

\begin{figure}
\plotone{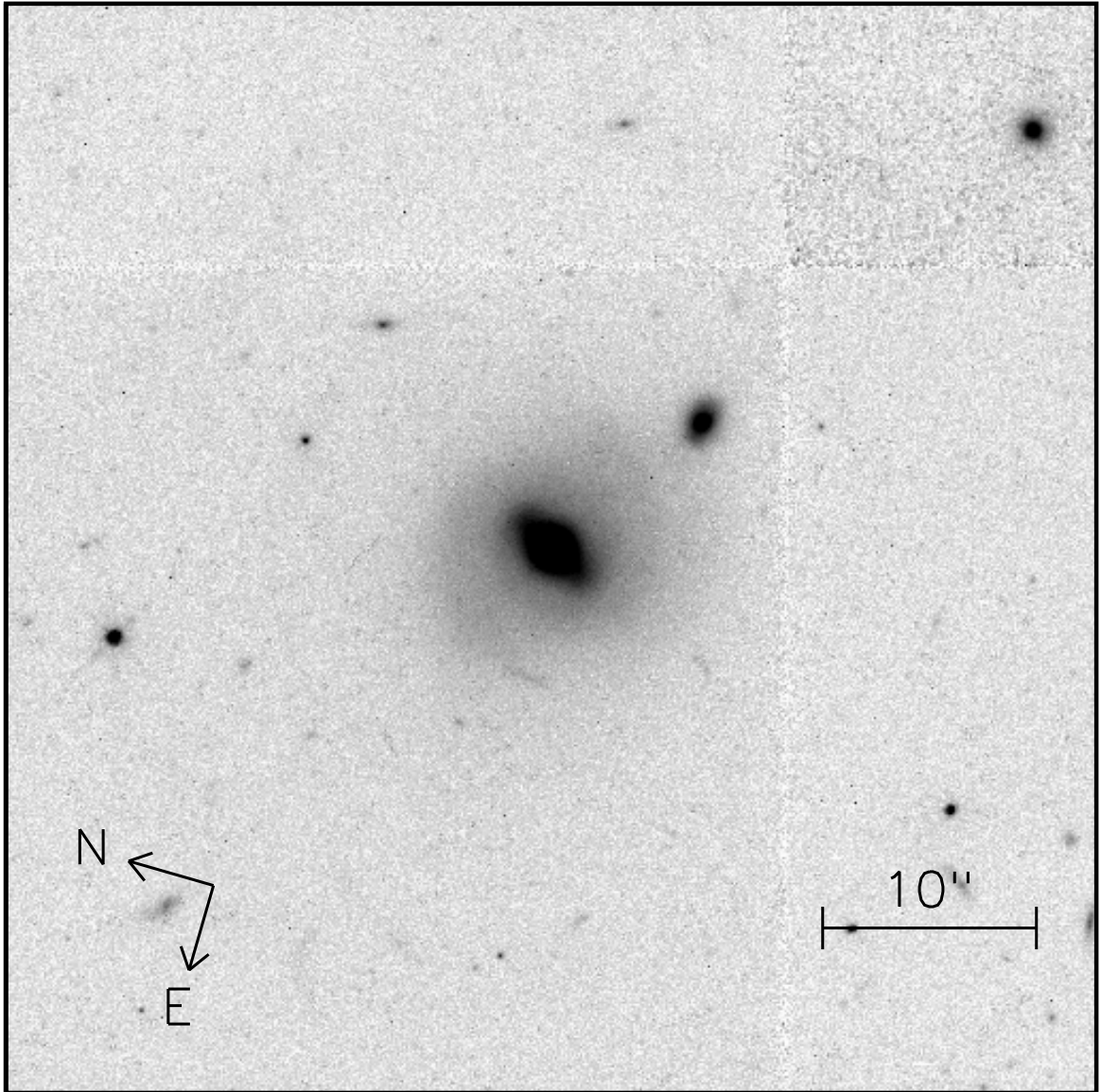}
\caption{Inner portion of the \hst\ WFPC2 F814W mosaic of Malin 1,
  displayed using a logarithmic stretch, showing the bulge, bar, and
  disk components.  The faint noisy vertical and horizontal lines are
  the boundaries between the WFPC2 CCDs. 
\label{fig-image}}
\end{figure}

The images were retrieved from the \hst\ data archive and individual
exposures were combined using standard tasks in
IRAF/STSDAS\footnote{IRAF is distributed by the National Optical
Astronomy Observatories, which are operated by the Association of
Universities for Research in Astronomy, Inc., under cooperative
agreement with the National Science Foundation.}.  Figure
\ref{fig-image} shows a portion of the F814W WFPC2 mosaic centered on
Malin 1.  An image of the entire WFPC2 mosaic is shown by
\citet{obi00}.  If the full \hst\ image is viewed with extremely high
contrast, some of the very low surface brightness outer spiral
structure can just barely be seen at radii of up to $\sim1\arcmin$
from the galaxy center, and the faint spiral features match the
structures visible in Fig. 2 of \citet{bot87}.  However, the S/N in
the WFPC2 image at these large radii is too low to perform useful
measurements of the surface brightness of the outer disk.  The F300W
image also has very low S/N and will not be discussed further in this
paper.

A new optical spectrum of Malin 1 was obtained at the Keck II
telescope on 2005 May 16 UT with the ESI spectrograph
\citep{sheinis02}.  The exposure time was 1800 s with a 0\farcs75-wide
slit, yielding a spectrum with $R \approx 6000$ over the range
3800--10000 \AA\ and S/N $\approx20$ per pixel in the continuum at
5100 \AA.  The slit was oriented at the parallactic angle for the
midpoint of the observation.  Spectral extraction was done using a
1\arcsec\ width, and the extracted echelle orders were
wavelength-calibrated using observations of HgNe, Xe, and CuAr lamps
and flux-calibrated using an observation of the standard star
BD~+28\arcdeg4211.  The ten individual echelle orders were combined
into a single spectrum, weighted by the S/N in the overlap regions
between orders.  Several K-giant stars were also observed on the same
night during twilight for use as velocity templates.

\section{Measurements and Results}

\subsection{Imaging}

The WFPC2 image clearly reveals the morphology of Malin 1 at small
angular scales where previous ground-based images were unable to
discern the details of its structure.  The galaxy is seen to have a
compact bulge dominating the light profile out to $r\approx1\arcsec$,
a bar of length $\sim6\arcsec$, and a nearly face-on disk with a hint
of spiral structure that can be traced out to $\sim6-7\arcsec$.  The
disk morphology appears smooth and there are no obvious dust lanes,
knots, or star-forming complexes.  Apart from the remarkable but
nearly invisible outer LSB structure, Malin 1 has the morphology of an
SB0/a galaxy.

To decompose the galaxy's structure into its subcomponents, model fits
to the WFPC2 image were performed using the 2-dimensional fitting
package GALFIT \citep{peng02}.  The components used in the model fit
included an exponential disk, a S\'ersic-law bulge, and a bar.  Since
GALFIT does not include bar models such as a Freeman (1966) bar or a
flat bar \citep{pri97}, the bar component was modeled using a S\'ersic
profile with the index $n$ constrained to be $\leq0.5$ to approximate
a bar with a flat core, and with this constraint the bar has $n = 0.5$
in the best-fitting model.  In addition, a central point source
component was added to allow for the possibility of unresolved
emission from the active nucleus.  The point-spread function for the
F814W filter was generated using the TinyTim software package
\citep{krist93}.

\begin{figure}
\begin{center}
\scalebox{0.3}{\includegraphics{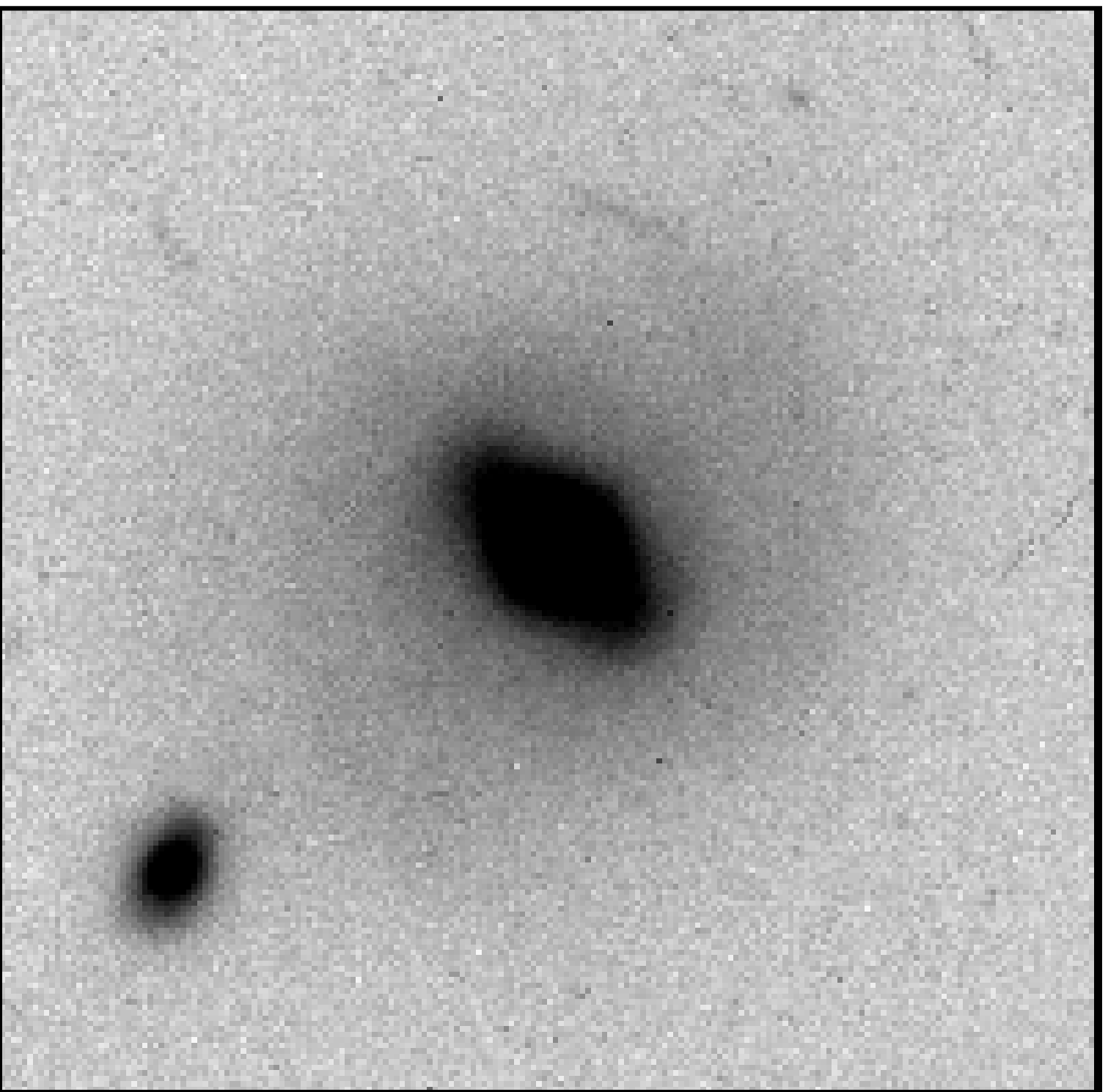}}
\hspace*{0.2in}
\scalebox{0.3}{\includegraphics{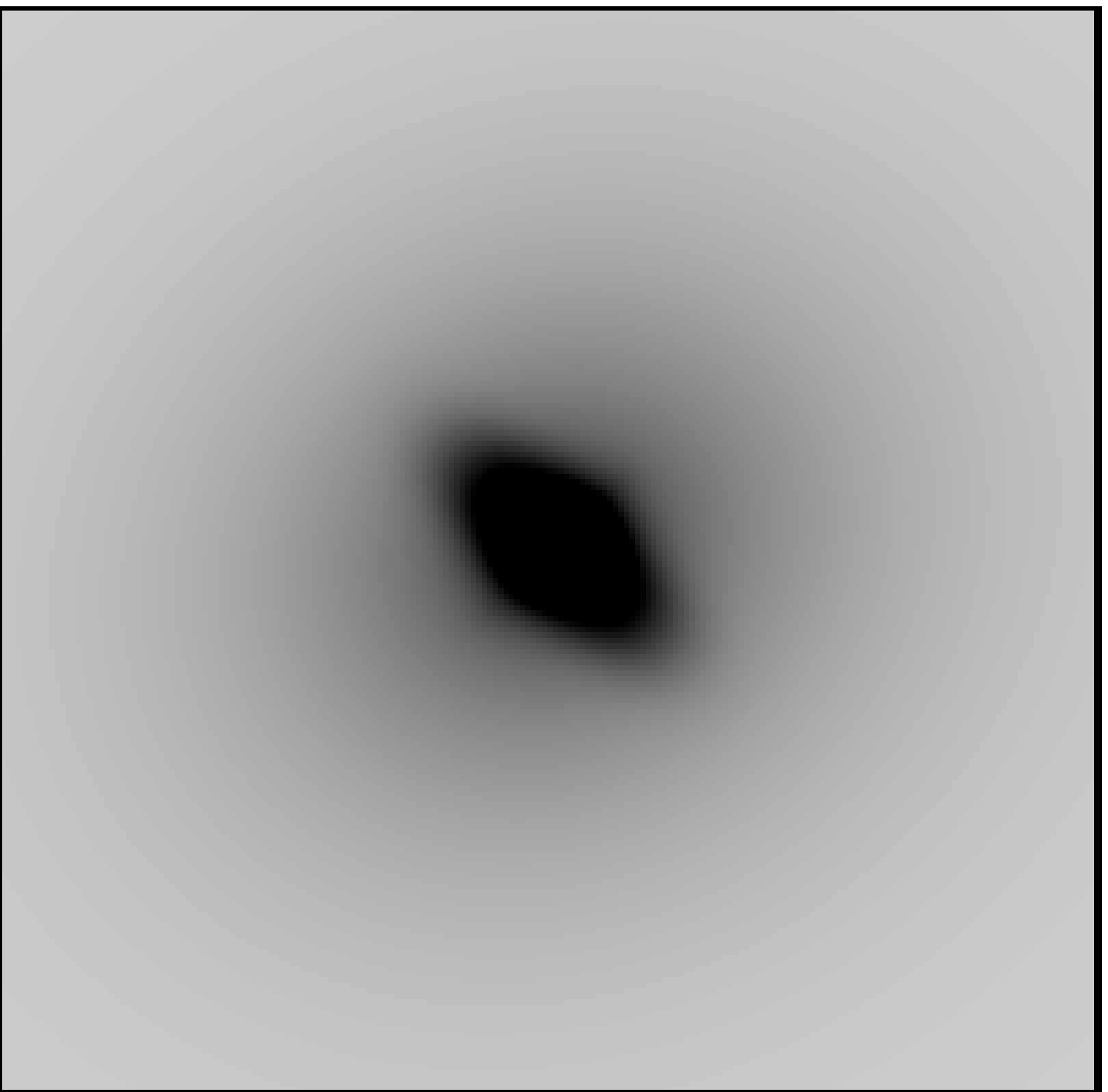}}
\hspace*{0.2in}
\scalebox{0.3}{\includegraphics{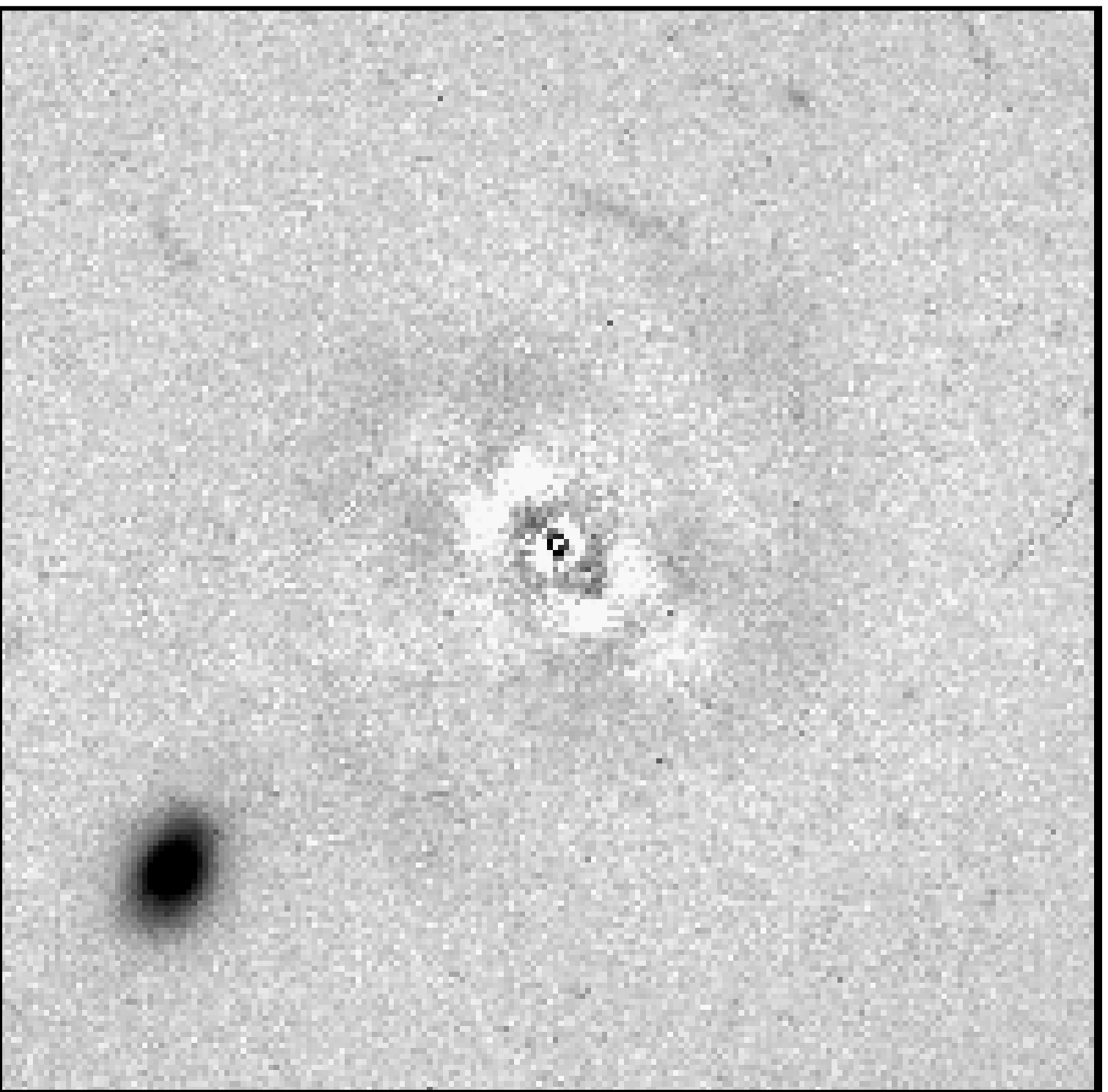}}
\end{center}
\caption{GALFIT decomposition of Malin 1.  \emph{Left panel:} A
  $20\arcsec\times20\arcsec$ portion of the WFPC2 image.  \emph{Middle
  panel:}  GALFIT model including bulge, bar, and disk components.
  \emph{Right panel:} Residuals after subtraction of galaxy model.
\label{galfit}}
\end{figure}

Figure \ref{galfit} shows the results of the GALFIT modeling.  The
structure of the galaxy is reproduced well overall, although there are
systematic residuals around the bulge and inner bar region.  These
residuals could be reduced somewhat if the constraint on the S\'ersic
index of the bar were relaxed, at the expense of having an unrealistic
bar profile with a very centrally peaked structure.  However, whether
the bar profile is constrained or not in the fit has almost no effect
on the central surface brightness or scale length of the disk
component, since the disk parameters are mainly determined from the
region around $r = 4\arcsec-6\arcsec$ where it dominates the galaxy
profile.  

\begin{figure}
\plotone{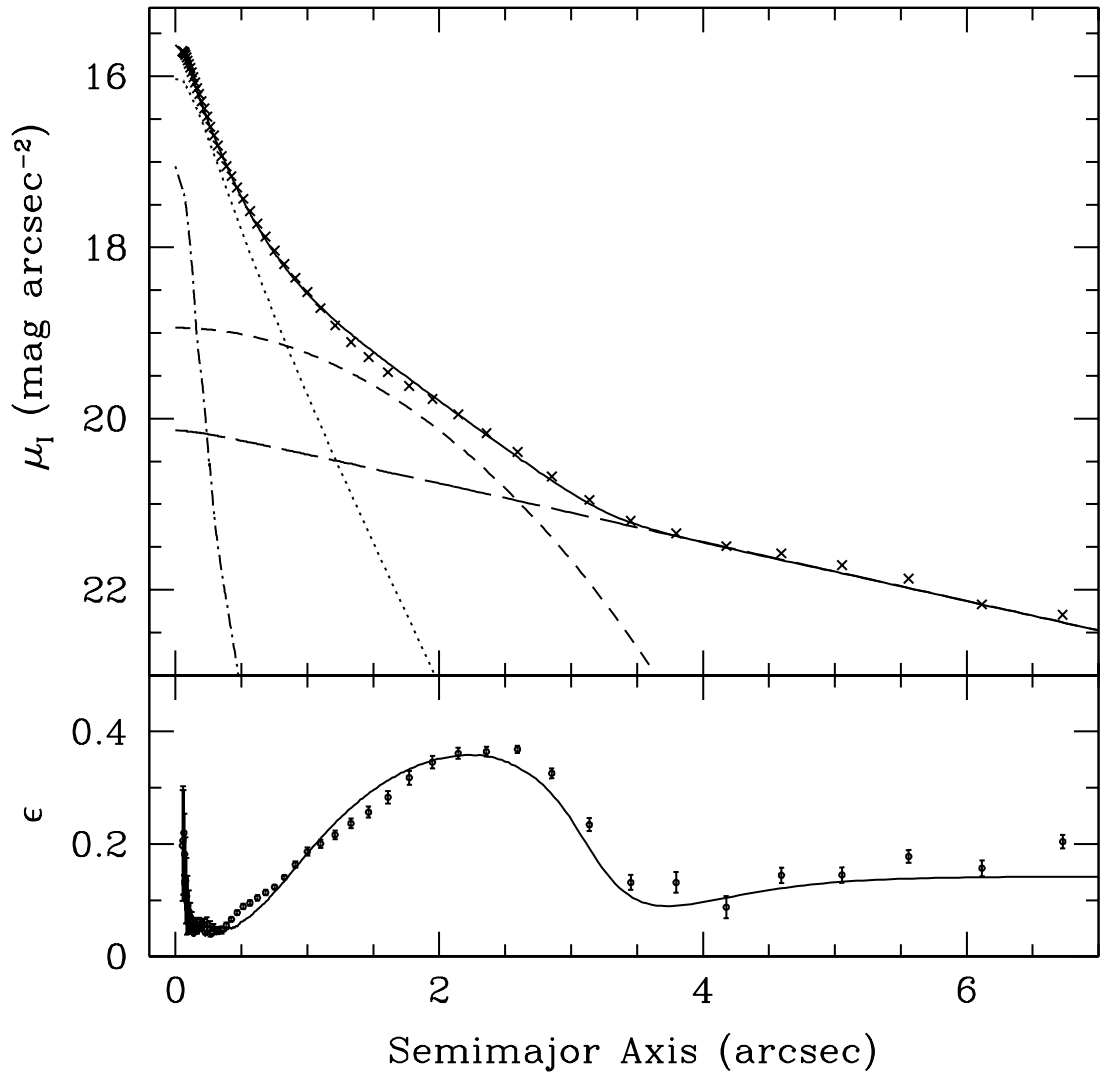}
\caption{$I$-band surface brightness profile for the inner region of
  Malin 1 (crosses), measured with the IRAF ELLIPSE task.  Model
  components are the central point source (dot-dashed curve), bulge
  (dotted), bar (short dashed), and disk (long-dashed), with the total
  galaxy model plotted as a solid line. The lower panel displays the
  ellipticity profile for the galaxy and for the GALFIT model.
\label{radprof}}
\end{figure}

\begin{figure}
\plotone{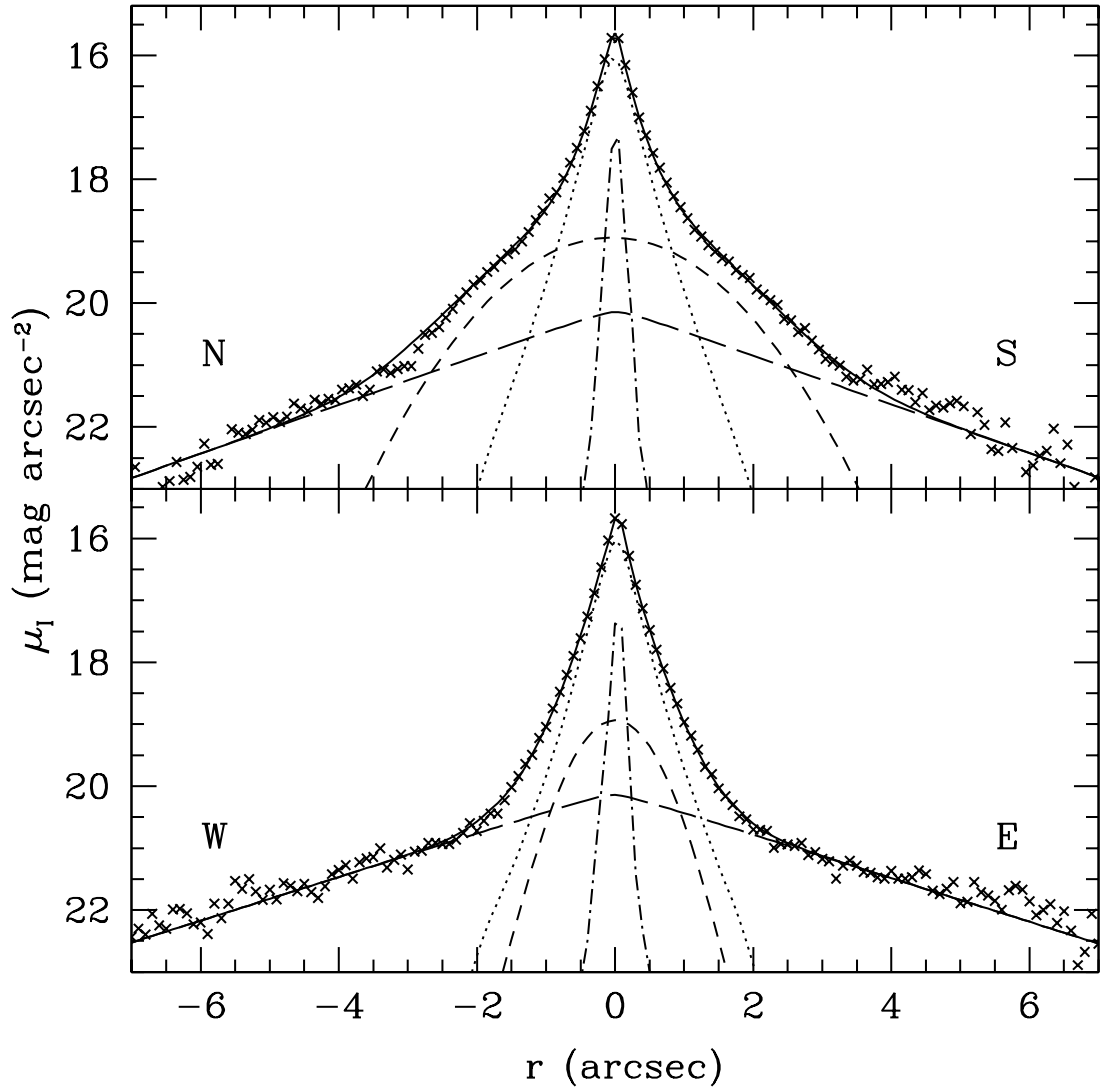}
\caption{Surface brightness cuts along the major and minor axes of the
  bar (upper and lower panels, repsectively).  Crosses represent
  measured surface brightness averaged over a 2 pixel-wide extraction
  through the image.   Model components are the central point source
  (dot-dashed curve), bulge (dotted), bar (short dashed), and disk
  (long-dashed), with the total galaxy model plotted as a solid line.
\label{cuts}}
\end{figure}

Figure \ref{radprof} shows the galaxy's radial profile as measured
from the WFPC2 image and from the GALFIT model.  The radial profiles
were measured using the IRAF task ELLIPSE, which fits elliptical
isophotes to the image at a series of fixed semimajor axis lengths,
following the methods described by \citet{jed87}.  Conversion from
\hst\ F814W filter magnitudes to standard Cousins $I$-band Vega-based
magnitudes was done using the SYNPHOT package in IRAF, assuming an
S0-type spectrum from the spectral atlas of \citet{kc96}.  The
brightness profiles include a correction for Galactic extinction
\citep[$A_I = 0.067$ mag;][]{sfd98}, a $K$-correction of 0.08 mag for
the $I$ band, determined using the \citet{kc96} S0 template spectrum,
and a correction for cosmological surface brightness dimming of 0.34
mag.  Brightness profiles derived from 0\farcs2-wide slices along the
major and minor axes of the bar are shown in Figure \ref{cuts}.  Aside
from small deviations in the disk due to the spiral arms, the major
and minor axis brightness profiles are very symmetric about the
nucleus.

Another issue to consider is the effect of inclination on the disk
surface brightness.  One simple method that is commonly used to
correct for inclination is to apply the relation
$\mu_\mathrm{corrected}$ = $\mu_\mathrm{observed} - 2.5C\log(a/b)$,
where $a$ and $b$ are the major and minor axis lengths of the disk,
and $C$ is a parameter whose value ranges from 0 for an optically
thick disk to 1 for the optically thin case \citep{dj96, sj98}.  From
the GALFIT decomposition, the disk axis ratio $b/a$ is 0.858,
consistent with the mean ellipticity of $\epsilon = 1-(b/a) = 0.14$
measured by the ELLIPSE routine over the disk-dominated region from $r
= 4$ to 6 arcsec.  This corresponds to a surface brightness correction
of $-0.17$ mag in the optically thin case.  Given the approximate
nature of this method and the small magnitude of the correction, we
choose not to apply this correction to the measured quantities, but we
consider it as a minor effect that would tend to increase the central
surface brightness by a small amount.

The structural parameters derived from the GALFIT modeling are listed
in Table 1.  While the formal fitting uncertainties from GALFIT are
negligibly small, the actual uncertainties on the derived model
parameters are predominantly systematic due to real deviations of the
galaxy components from the simple fitting functions used by GALFIT and
are therefore difficult to estimate.  Trial GALFIT runs with slightly
different fitting models (i.e., without constraints on the bar
S\'ersic index $n$ or without including a point source component)
yielded magnitude differences of order $\sim0.2$ mag for the bulge and
bar components relative to the best-fit model listed in Table 1, which
gives some indication of the likely systematic uncertainties in the
decomposition.  Different choices of bulge and bar models or omission
of the nuclear point source led to relatively small changes in the
exponential disk parameters, at the level of $<0.1$ mag in total
magnitude and $<5\%$ in scale length.

\subsection{Spectroscopy}

As previously shown by \citet{ib89}, the bulge spectrum is consistent
with a predominantly old stellar population.  The stellar velocity
dispersion of the bulge was measured from the Keck spectrum by direct
fitting of broadened and diluted stellar spectra in the wavelength
domain, following the methods described by \citet{bhs02}.  The fit was
performed over the rest wavelength range 5210--5480 \AA, which
contains Fe $\lambda$5270 and several other reasonably strong
features.  The \mgb\ lines were excluded from the fit since the galaxy
spectrum has a lower Fe/$\alpha$ abundance ratio than typical nearby K
giant stars, as is generally found for high-dispersion galaxies.  Weak
[\ion{N}{1}] $\lambda5200$ emission appears to be present as well.
Fits were performed using nine template stars with spectral types
between G8 III and K3 III were.  The best fit was found with a K0
star, giving $\sigma = 196 \pm 15$ \kms, where the final uncertainty
is the sum in quadrature of the fitting uncertainty from the
best-fitting template (10 \kms) and the standard deviation of the
results from fitting all nine templates (11 \kms). Figure
\ref{figdispersion} shows the fit of the broadened KO star to the
spectrum of Malin 1 over this region.

\begin{figure}
\plotone{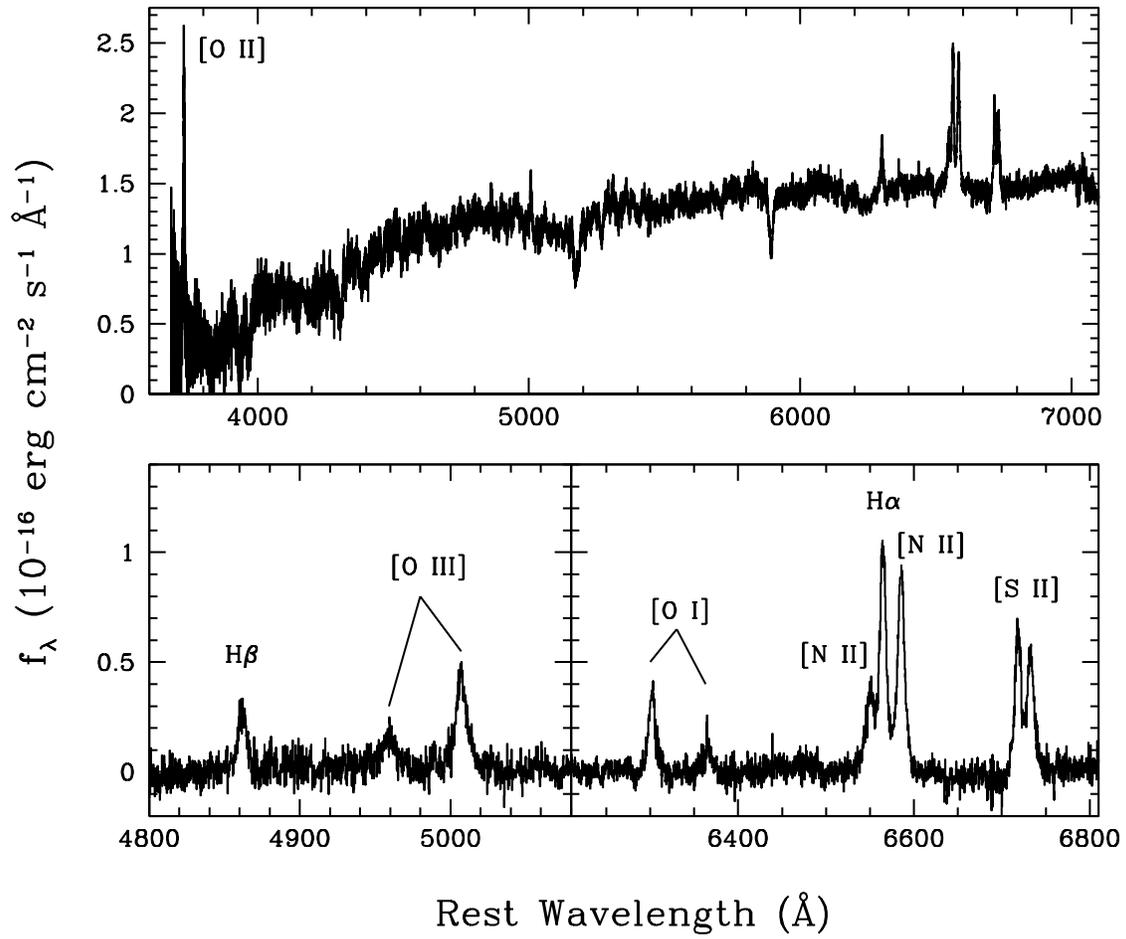}
\caption{Keck ESI spectrum of Malin 1.  \emph{Upper panel:}  A portion
  of the complete ESI spectrum.  \emph{Lower panels:}  The
  starlight-subtracted spectrum in the regions around the \hbeta\ and
  \hal\ emission lines.
\label{figspectrum}
}
\end{figure}

\begin{figure}
\plotone{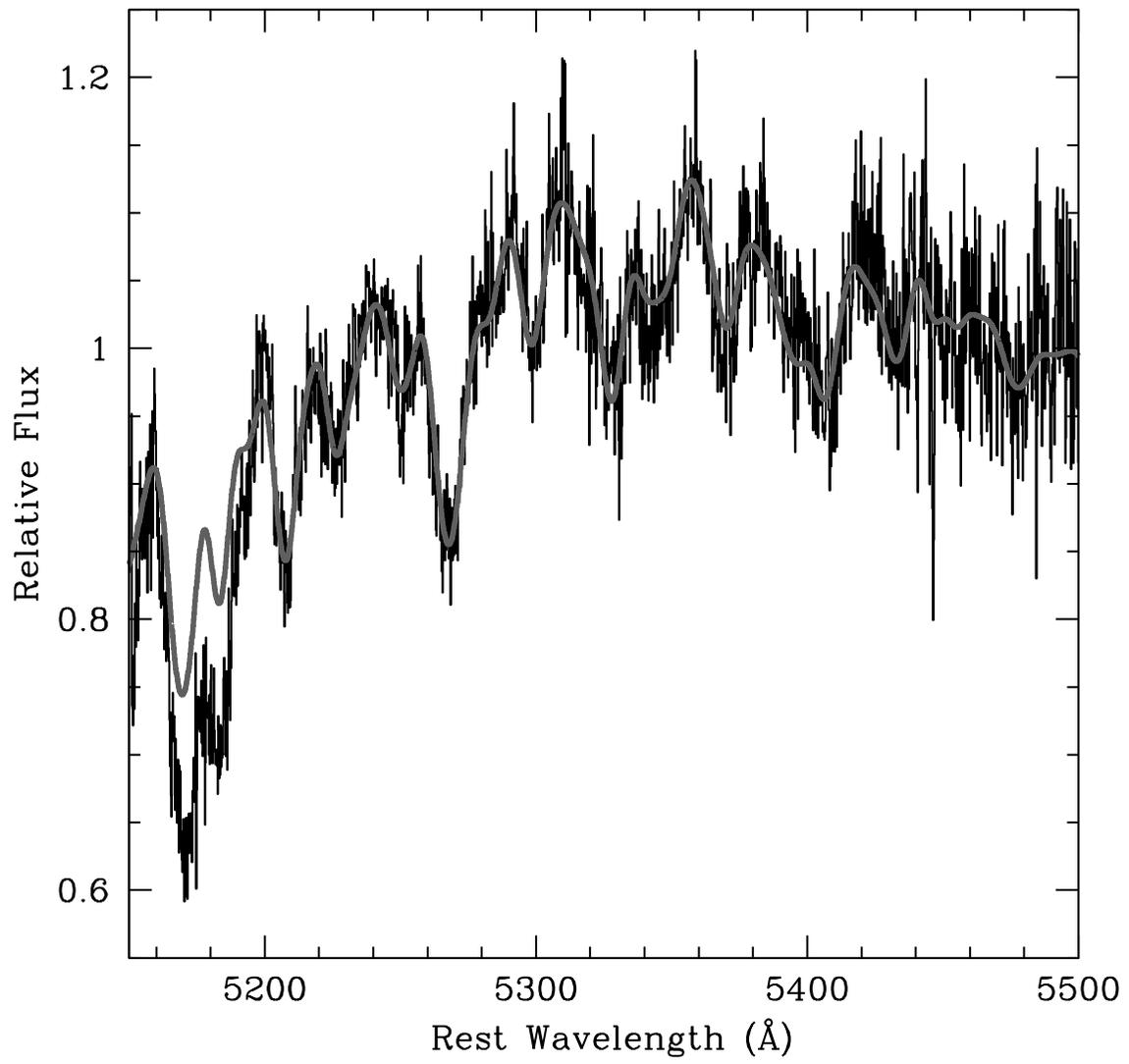}
\caption{Best fit of a broadened K-giant template star, HD 92068 (K0
  III), to the spectrum of Malin 1.
\label{figdispersion}
}
\end{figure}

The velocity dispersion fitting code was also used to generate
starlight-subtracted spectra in the regions surrounding the \hbeta\
and \hal\ emission lines.  Using the best-fitting K-type stellar
template and allowing for a featureless continuum dilution yielded an
adequate continuum subtraction over these regions, giving a pure
emission-line spectrum (Figure \ref{figspectrum}).

Reddening-corrected emission line flux ratios measured from the
nuclear spectrum are [\ion{O}{2}] $\lambda3727$ / [\ion{O}{3}]
$\lambda5007$ = 3.1, [\ion{O}{3}] $\lambda5007$ / \hbeta\ = 1.9,
[\ion{O}{1}] $\lambda6300$ / \hal\ = 0.3, and [\ion{N}{2}]
$\lambda6584$ / \hal\ = 0.85.  These measurements are consistent with
a LINER classification \citep{h80, hfs97a}.  \citet{ib89} had
previously found [\ion{O}{3}] / \hbeta\ = 4.8 and classified Malin 1
as a high-excitation Seyfert galaxy; presumably this resulted from not
having performed a starlight subtraction and therefore missing most of
the \hbeta\ flux.  Approximately 30\% of S0--Sa galaxies contain LINER
nuclei \citep{hfs97b} so Malin 1 is not unusual in exhibiting this
type of activity.  There is at best weak evidence for a broad
component to the \hal\ emission line.  Model fits including 3 Gaussian
components for the three narrow emission lines are somewhat improved
by the addition of a broad component (having a best-fitting width of
FWHM = $2010\pm80$ \kms), but the relatively low S/N of the spectrum
precludes any definitive conclusions regarding the possible presence
of a broad-line component.

\section{Discussion}


The measurements described above indicate a very different structure
for Malin 1 than that which has previously been found.  \citet{bot87}
fit the galaxy profile with a model consisting of an $r^{1/4}$-law
bulge and an exponential disk, and determined that the bulge had $r_e
= 2\farcs9\pm0\farcs5$, which corresponds to 4.4 kpc for the cosmology
assumed in this paper.  \citet{pic97} found similar results from newer
ground-based imaging data.  The \hst\ image shows that the bulge
radius is smaller than this value by nearly an order of magnitude, and
it seems likely that the bulge region described by \citet{bot87} was
actually the combined light from the bulge, bar, and normal disk of
the galaxy.  Since the surface brightness of the disk within several
arcseconds of the nucleus is much fainter than the surface brightness
of the bulge or bar, it is not surprising that it would be difficult
to recognize the disk in ground-based imaging of average seeing.
Interestingly, \citet{bot87} note that the nucleus ``appears to be
surrounded by extended nebulosity of somewhat high surface
brightness''.  In retrospect, this ``nebulosity'' must have been
starlight from the disk itself.

Recent work by \citet{agu05} on the $I$-band photometric properties of
SB0 galaxies provides an ideal comparison sample to examine the
structure of Malin 1 in the context of galaxies of similar Hubble
type.  Based on a sample of 14 nearby galaxies, they found that SB0
galaxies generally have nearly exponential bulge profiles with $n =
1.48 \pm 0.16$ and $r_e$ in the range 0.3 to 1.0 kpc with a mean of
0.6 kpc, so Malin 1's bulge properties are typical for its Hubble
type. With a bulge absolute magnitude of $M_I = -20.9$ mag and $\sigma
= 196$ \kms, Malin 1 also falls within the normal range of SB0
galaxies in the Faber-Jackson relation (see Fig. 5 of Aguerri \etal\
2005).

\begin{figure}
\plotone{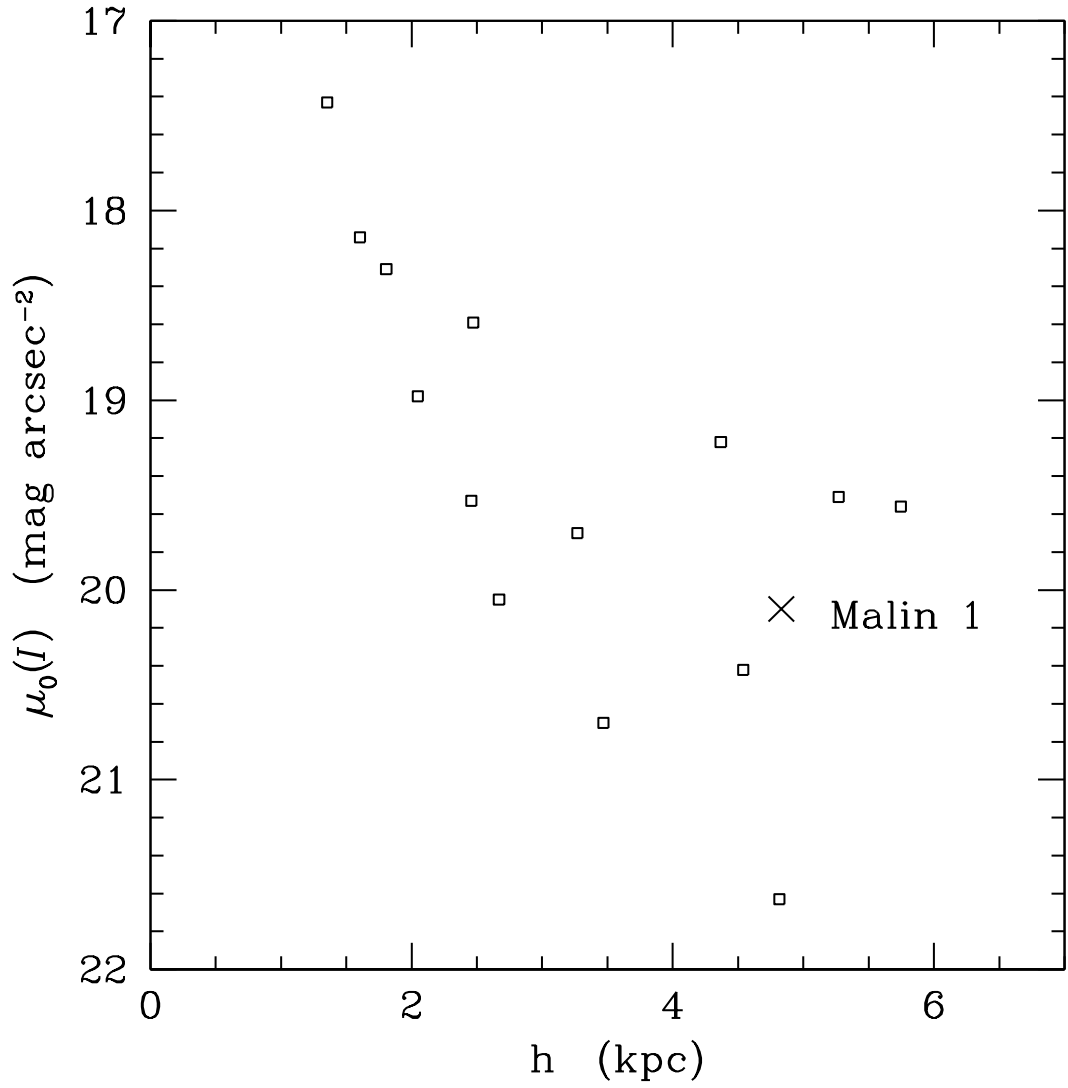}
\caption{$I$-band central surface brightness vs. scale length for disk
  components in SB0 galaxies.  Open squares are from \citet{agu05}.
  Malin 1 is represented by an $\times$ symbol.
\label{sb0}}
\end{figure}

The disk of Malin 1 also has properties similar to the SB0 sample.
Figure \ref{sb0} plots the central $I$-band surface brightness against
disk scale length.  While Malin 1 falls toward the lower end of the
sample in terms of its surface brightness, it is still consistent with
the normal range for this Hubble type.  The only aspect of Malin 1's
structure that appears different from the SB0 sample is the ratio of
bulge to disk radius. \citet{agu05} find a surprisingly tight coupling
with $r_e/h = 0.20\pm0.01$ for their sample, while Malin 1 has $r_e/h
= 0.13$.  A portion of this difference could be the result of the
different bar model adopted for the GALFIT analysis, however, since
\citet{agu05} used Freeman bar or flat bar models in their
1-dimensional radial profile fits, and different assumptions for the
bar profile will directly affect the flux inferred to arise from the
outer part of the bulge.  In any case, a bulge-to-disk scale length
ratio of 0.13 is well within the normal range for low or high surface
brightness spiral galaxies of a range of Hubble types \citep{dj96,
sj98, bei99}.  Overall, these results indicate that out to $r \approx
10$ kpc, the optical structure of Malin 1 does not deviate very
significantly from the general population of early-type barred
galaxies.

The low-surface brightness structure at large radii appears to be a
photometrically distinct component of the galaxy's structure rather
than a smooth extension of the normal inner disk.  For a disk scale
length of 4.8 kpc, the extrapolated surface brightness of the inner
disk at $r = 50\arcsec$ (equivalent to $\sim75$ kpc) would be
undetectably faint: 36.1 mag arcsec\persq\ in the $B$ band, which is
far fainter than the actual outer disk brightness of $\mu_V \approx
27$ mag arcsec\persq\ at this radius \citep{bot87}.  There may be a
break in the exponential light profile of the disk at radii
greater than $\sim7\arcsec$, but deeper images would be needed to
trace the surface brightness profile in the transition region between
the inner disk and the extended outer disk.


Given the presence of this newly identified disk in Malin 1, should it
still be considered an LSB galaxy?  LSB galaxies are generally
classified as such based on the extrapolated central surface
brightness of their disk components.  A crucial point is that the LSB
classification applies only to the disk component.  Giant LSB disk
galaxies have high surface brightness bulges that dominate the light
of the central regions \citep[e.g.,][]{spr95}.  Thus, the \emph{total}
light profiles of giant LSB galaxies apparently always have high
surface brightness central regions, and a radial profile decomposition
is required in order to determine whether a galaxy should be
classified as having a giant LSB disk or not.  Clearly, to perform
such classifications in a meaningful way, observations with sufficient
spatial resolution to decompose the bulge and inner disk components
accurately are required.

While there is perhaps no universal agreement on the surface
brightness threshold for a galaxy disk to be considered an LSB disk, a
common criterion is a central surface brightness fainter than $\mu_B =
23.0$ mag arcsec\persq\ \citep{ib97}.  As noted by \citet{bim97}, a
galaxy with a central surface brightness fainter than this level would
represent a $>4\sigma$ deviation from the distribution of surface
brightness found by \citet{fre70} and would therefore be a very
unusual outlier if the actual distribution of disk surface
brightnesses were described by Freeman's law.  To compare the disk
properties of Malin 1 with this criterion, the $B$ magnitude of the
Malin 1 disk was estimated by performing synthetic photometry on the
S0 template spectrum from \citet{kc96}.  The $(B-I)$ color index for
this spectrum is 2.2 mag.  Assuming this $(B-I)$ color for the disk of
Malin 1, it has an estimated $\mu_0(B) = 22.3$ mag arcsec\persq, which
would not qualify it as an LSB disk\footnote{For comparison,
\citet{fuk95} find an average $(B-I)$ color of 2.0 mag for S0
galaxies; using this value instead of the SYNPHOT color would yield
$\mu_0(B) = 22.1$ mag arcsec\persq.}.  In the classification system of
\citet{mg96}, Malin 1 would have an ``intermediate surface
brightness'' disk.

Another related criterion that has been applied to distinguish low and
high surface brightness galaxies is a ``diffuseness index'' for the
disk, which combines central surface brightness and scale length into
a single parameter.  The diffuseness index criterion given by
\citet{spr95} is $\mu_B(0) + 5\log(h) > 27.0$ for a galaxy to be
classified as a giant LSB disk.  Based on the GALFIT results and the
assumed $(B-I)$ color index of 2.2 mag, the disk in Malin 1 has
$\mu_B(0) + 5\log(h) = 25.0$, which lies well inside the high surface
brightness range for this parameter.  Thus, based on either central
surface brightness or diffuseness, the disk of Malin 1 out to $r
\approx 10$ kpc should \emph{not} be considered an LSB disk according
to conventional criteria.

Malin 1 still might be considered an LSB galaxy based on the average
surface brightness over its \emph{entire} disk area, however.
According to \citet{pic97}, the galaxy's total luminosity is $M_V =
-22.9\pm0.4$ mag.  The \hst\ measurements yield $M_I = -22.4$ mag for
the combined bulge, bar, and normal disk components.  Assuming $(V-I)
= 1.3$ mag for an S0 galaxy as determined from SYNPHOT, this
corresponds to $M_V = -21.1$ mag.  Thus, the comparison between the
\hst\ and ground-based measurements indicates that the majority of the
galaxy's total optical luminosity lies in the extended disk region
beyond the normal disk.  It would be useful to confirm this result
with new ground-based imaging data and a single profile decomposition
over the entire radial range of the galaxy; the \hst\ image could be
used to exclude background galaxies and improve the measurement
accuracy for the outer disk brightness profile.

In comparison with all other known disk galaxies, Malin 1 has always
appeared as a unique and distant outlier in plots of central surface
brightness against disk scale length \citep[see, e.g.,][]{bot87,
spr95, bim97, dss97, vdb98}.  The previously measured disk scale
length and central surface brightness ($h \approx 70$ kpc; $\mu_0(V) =
25.5$ mag arcsec\persq) from \citet{bot87} are far outside the ranges
found for normal spirals or even for other giant LSB galaxies.  In the
giant LSB sample of \citet{spr95}, for example, the disk scale lengths
range from 5 to 13 kpc and the faintest disk has $\mu_B(0) = 24.2$ mag
arcsec\persq.  Clearly, however, when the \emph{normal} disk of Malin
1 is compared with the disk properties of other galaxies in these
diagrams, it is no longer an outlier.  It might be argued that the
extended outer structure in Malin 1 still has the form of a disk with
a very low extrapolated central surface brightness and large scale
length and that this extended disk remains an outlier relative to all
other known galaxies.  However, it seems more reasonable and
consistent to compare the normal, central disk of Malin 1 with those
of other galaxies, and to consider its extended outer disk in the
context of the extreme outer disks of other nearby spiral galaxies,
which sometimes display structures similar to (albeit less extended
than) that of Malin 1.

Recent work has revealed a great deal of detail and a diversity of
properties among the extreme outer disk regions of some nearby spiral
galaxies.  Deep narrow-band observations of some late-type spirals
have detected \ion{H}{2} regions far beyond the galaxies' optical
radii \citep{fer98b, lr00}, and \emph{GALEX} observations of the
nearby spiral M83 have detected UV-bright stars in the outer disk at
distances of greater than 20 kpc from the galaxy center \citep{thi05}.
In these examples, the regions of recent star formation have
filamentary, spiral-like structures with an appearance similar to the
outer disk regions of Malin 1.  \citet{fer98a} noted that the distant
outer regions of spiral disks have physical properties (such as low
column densities of gas and high gas mass to stellar mass ratios) that
are very similar to the inferred properties of Malin 1.  It is now
apparent that this similarity is not merely a coincidence-- the
parameters previously found for Malin 1 were precisely those of its
extreme outer disk region.  Extended outer disk structure has recently
been found out to $r \approx 40$ kpc in M31 as well \citep{fer02,
iba05}, but in the case of M31 the extended outer structure appears to
be a smooth continuation of the exponential profile of the main disk,
in contrast to the photometrically decoupled outer structure of Malin
1.  The stellar disk of the late-type spiral NGC 300 also follows a
single exponential profile from its inner regions to the outer disk at
distances up to 10 scale lengths \citep{bh05}.

New observations of early-type barred galaxies have revealed
surprising outer disk structures as well.  \citet{ebp05} have found
that at least 25\% of SB0-SBb galaxies have photometric profiles
characterized by an outer exponential disk component with a larger
scale length than the inner, main disk component.  The transition from
the inner to outer disk occurs at 3--6 scale lengths of the inner disk
and at surface brightnesses of 22.6--25.6 mag arcsec\persq, and spiral
structure is often seen in the outer disk region.  Erwin \etal\ call
such structures ``antitruncated'' disks.  Malin 1 appears to fit
within this category, although deeper images would be needed to
determine the structure of the transition region between the inner and
outer disks.

Simulations by \citet{pmb06} have explored one possible mechanism that
could form such highly extended outer disks, by the tidal shredding of
dwarf galaxies in the outskirts of an M31-sized galaxy.  With an
initial apocenter of $r=75$ kpc, the dwarf satellites were tidally
distorted into extended disks having spiral-like filamentary
structure, extending to over 100 kpc from the center of the larger
galaxy.  The tidal debris tends to evolve into a nearly exponential
profile, with scale lengths of up to 50 kpc, depending on the
compactness of the original stellar distribution in the dwarf galaxy.

Could other giant LSB galaxies possess inner, high surface brightness
disks that have gone undetected?  Several of the other known giant LSB
galaxies lie at distances comparable to that of Malin 1
\citep[e.g.,][]{spr95}, and inadequate spatial resolution could cause
a normal disk to be missed in a bulge-disk decomposition.  This would
be a natural pitfall of fitting a simple 2-component model to a galaxy
with a bright central region (including a bulge and disk) and a very
diffuse and extended outer disk that was brighter than a simple
extension of the inner disk profile at large radii. The outer disk
would dominate the fit at large radii and therefore determine the
scale length and central surface brightness of the exponential disk
component in the fit, and the inner, high surface brightness disk, if
present, would be subsumed in the outer part of the bulge profile.
Deep, high-resolution imaging of a representative sample of giant LSB
galaxies would be useful to clarify the nature of these objects.

\section{Conclusions}

The stellar disk of Malin 1 has long been thought to be an extreme and
distant outlier compared to the disk properties of all other known
spiral galaxies.  The archival \hst\ data demonstrates that the
central structure of Malin 1 is that of a normal early-type barred
spiral galaxy, and the central surface brightness of its disk is
$\sim4$ magnitudes brighter than the value originally found by
\citet{bot87}.  Thus, it seems reasonable to consider Malin 1 not as
an LSB galaxy, but rather as a galaxy with the normal structural
components of an early-type barred spiral, which is embedded in a
remarkably extended, optically faint, and gas-rich outer structure
beyond its normal disk.

\citet{bim97} remark that giant LSB galaxies such as Malin 1 are
``truly enigmatic in that 'normal' formation processes were at work to
create the bulge component but no conspicuous stellar disk ever formed
around this bulge.''  Malin 1 is perhaps a somewhat less enigmatic
galaxy once its normal disk has been recognized.

\acknowledgments

I am grateful to Joe Shields and Karen O'Neil for discussions that
motivated this work and to Luis Ho, Marc Seigar, Peter Erwin, Chien
Peng, and the anonymous referee for many helpful suggestions.  Support
for the Keck observations was generously provided by the UC Irvine
Physical Sciences Innovation Fund.  The author wishes to recognize and
acknowledge the very significant cultural role and reverence that the
summit of Mauna Kea has always had within the indigenous Hawaiian
community.  We are most fortunate to have the opportunity to conduct
observations from this mountain.

\clearpage

\begin{deluxetable}{lc}
\tablewidth{3in}
\tablecaption{Structural Properties of Malin 1.}
\tablehead{\colhead{Parameter} & \colhead{Value}}
\startdata
Bulge $r_e$ & 0.6 kpc \\
Bulge S\'ersic index $n$ & 1.24 \\
Bulge $I$ magnitude &  17.0 mag \\
Bar $I$ magnitude & 17.6 mag \\
Point source $I$ magnitude & 20.6 mag \\
Disk $\mu_0(I)$ & 20.1 mag arcsec\persq\ \\
Disk scale length $h$ & 4.8 kpc \\
Bulge velocity dispersion & $196 \pm 15$ \kms 
\enddata

\tablecomments{Properties listed here are determined from the GALFIT
  decomposition of the WFPC2 image and from the Keck spectrum.  The
  magnitudes listed above include a correction for Galactic extinction
  of 0.067 mag.  Additionally, a $K$-correction of 0.08 mag and a
  correction for cosmological surface brightness dimming have been
  applied to the value of the disk central surface brightness
  $\mu_0(I)$. }
  
\end{deluxetable}

\clearpage

\clearpage

\clearpage

\clearpage

\clearpage

\clearpage

\clearpage

\clearpage

\end{document}